# Enhanced Ad-Hoc on Demand Multipath Distance Vector Routing Potocol (*EAOMDV*)

Mrs.Sujata V.Mallapur
Department of Information Science and Engineering
Appa Institute of Engineering and Technology
Gulbarga, India
Sujatank000@yahoo.co.in

Prof. Sujata .Terdal
Department of Computer Science and Engineering
P.D.A College of Engineering
Gulbarga, India
Suja_pst@rediffmail.com

*Abstract*—Due to mobility in Ad-Hoc network the topology of the network may change randomly, rapidly and unexpectedly, because of these aspects, the routes in the network often disappear and new to arise. To avoid frequent route discovery and route failure EAOMDV was proposed based on existing routing protocol AOMDV. The EAOMDV (Enhanced Ad-Hoc on Demand Multipath Distance Vector) Routing protocol was proposed to solve the "route failure" problem in AOMDV. EAOMDV protocol reduces the route failure problem by pre-emptively predicting the link failure by the signal power received by the receiver (pr). This proposed protocol controls overhead, increases throughput and reduces the delay. The EAOMDV protocol was implemented on NS-2 and evaluation results show that the EAOMDV outperformed AOMDV.

I. INTRODUCTION

A mobile Ad-Hoc Network (MANET) [1] is a collection of mobile nodes that form a wireless network without the use of any fixed base station. Each node acting as both a host and a router arbitrarily and communicates with others via multiple wireless links, therefore the network topology changes greatly. The routing protocols proposed so far can be divided in to 2 categories: proactive routing protocol and reactive routing protocol. Reactive routing protocol, which initiates route computation only on demand, performs better than proactive routing protocol, which always maintains route to destination by periodically updating, due to its control overhead.
In such dynamic network, it is an essential to get route in time, perform the routing with maximal throughput and minimal control overhead. Several on-Demand routing protocol have been proposed. In such protocols, nodes build and maintain the routes as they are needed. Examples of these protocols include the Dynamic Source Routing (DSR) [7] and Ad hoc On-Demand Distance Vector AODV Routing protocol [2]. These protocols initiate a route discovery process whenever a node needs route to a particular destination. In AODV the source broadcasts a route Request (RREQ) packet in the network to search for route to the destination. When a RREQ reaches either the destination or an intermediate node that knows a route to the destination, a Route Reply (RREP) packet is unicast back to the source. This establishes a path between the source and destination. Data is transferred along this path until one of the links in the path breaks due to node mobility. The source is informed of this link failure by means of a Route Error (RERR) packet from the node upstream of the failed link. The source node then re-initiates a route discovery process to find a new route to the destination. It is a single path routing, which needs new route discovery whenever path breaks. To overcome such inefficiency, several studies have been presented to compute multiple paths. If primary path breaks, they provide alternative paths to send data packets without executing a new discovery.

A mobile Ad-Hoc Network (MANET) [1] is a collection of mobile nodes that form a wireless network without the use of any fixed base station. Each node acting as both a host and a router arbitrarily and communicates with others via multiple wireless links, therefore the network topology changes greatly. The routing protocols proposed so far can be divided in to 2 categories: proactive routing protocol and reactive routing protocol. Reactive routing protocol, which initiates route computation only on demand, performs better than proactive routing protocol, which always maintains route to destination by periodically updating, due to its control overhead.
In such dynamic network, it is an essential to get route in time, perform the routing with maximal throughput and minimal control overhead. Several on-Demand routing protocol have been proposed. In such protocols, nodes build and maintain the routes as they are needed. Examples of these protocols include the Dynamic Source Routing (DSR) [7] and Ad hoc On-Demand Distance Vector AODV Routing protocol [2]. These protocols initiate a route discovery process whenever a node needs route to a particular destination. In AODV the source broadcasts a route Request (RREQ) packet in the network to search for route to the destination. When a RREQ reaches either the destination or an intermediate node that knows a route to the destination, a Route Reply (RREP) packet is unicast back to the source. This establishes a path between the source and destination. Data is transferred along this path until one of the links in the path breaks due to node mobility. The source is informed of this link failure by means of a Route Error (RERR) packet from the node upstream of the failed link. The source node then re-initiates a route discovery process to find a new route to the destination. It is a single path routing, which needs new route discovery whenever path breaks. To overcome such inefficiency, several studies have been presented to compute multiple paths. If primary path breaks, they provide alternative paths to send data packets without executing a new discovery.

The current multipath routing protocols finds multiple route during Route Discovery process [11]. The best path that is with the shortest hop count is chosen as the primary path for data transfer while other paths are used only when primary path fails. These protocols do not perform any prediction of route failure before the path breaks. As a result it leads the problem of frequent route discovery for data transmission.

To remedy this problem this paper presents a new routing protocol called the *Enhanced Ad-hoc on Demand Multipath Distance Vector (EAOMDV)* routing protocol. This protocol is proposed based on the AOMDV [5] to provide stable path in Ad-Hoc network. EAOMDV is an extension of AOMDV. EAOMDV uses the Node-Disjoint Multipath routing Protocol. The goal of EAOMDV is to provide efficient recovery from "route failure" problem in dynamic network. To achieve this goal it computes "the received power of each receiver node" using "link failure prediction technique" to discover the route from source and destination. A source node sends the RREQ packet, if the receiver node has less received power than threshold then drops the packet and sends the warning packet to the sender than a source node not select the route containing, therefore, the selected routing path exclude the nodes that are going out of threshold. A Route Maintenance mechanism is implemented to repair a broken route, finds the new route when all route failed.

The rest of the paper is organized as follows. Section II describes related work. Section III describes the AOMDV route discovery and route maintenance. The proposed protocol is presented in section IV, and its performance is evaluated and compared with that of AOMDV in section V. Some conclusions are given in section VI.

## II. RELATED WORK

This section summarizes various examples of on-demand multipath routing protocols especially from the viewpoint of route discovery strategy

AODV Backup Routing (AODV-BR) [3] enhances the AODV by letting each neighboring node of a primary route maintain its own backup route. When the node over a primary route detects a link failure, it transmits a route update message and a neighboring node receiving this message activates the backup route. A problem of this approach is limitation of the route selection that is at most within one hop distance.

AOMDV (Ad hoc On-demand Multipath Distance 'Vector routing) [5] is a sophisticated protocol which produces multiple routes with loop-free and link-disjoint properties. When an intermediate node receives copies of a RREQ packet, it compares a hop count field in a packet with the minimum hop count, called *advertised-hopcount,* stored in a routing table for previous RREQ packets. Only a packet with the minimum hop count is accepted to avoid routing loops.

Furthermore, a *firsthop* field in a RREQ packet is then compared with the firsthop-list in a routing table. When a route with node-disjoint property (new *firsthop)* is found, a new reverse route is recorded. Destination returns RREP packets accordingly, and multiple routes with link-disjoint' property are established at a source node. A problem of AOMDV is that several efficient routes may be missed due to strong restriction by the firsthop field. Another problem is lack of backup route maintenance that causes timeout expiration of backup routes.

## III. AOMDV OVERVIEW

Ad Hoc On-Demand Multipath Distance Vector Routing Protocol [5] is one of the most used Ad-Hoc routing protocol. It is a reactive routing protocol based on DSDV. AOMDV is designed for networks with tens to thousands of mobile nodes. The main idea in AOMDV is to compute multiple paths during route discovery. It is designed primarily for highly dynamic ad hoc networks where link failures and route breaks occur frequently. When single path on-demand routing protocol such as AODV is used in such networks, a new route discovery is needed in response to every route break. Each route discovery is associated with high overhead and latency. This inefficiency can be avoided by having multiple redundant paths available.

The AOMDV protocol has two main components:
1. A route update rule to establish and maintain *multiple loop-free* paths at each node.
2. A distributed protocol to find *node-disjoint* paths that is route discovery.

In AOMDV a new route discovery is needed only when all paths to the destination break. A main feature of the AOMDV protocol is the use of routing information already available in the underlying AODV protocol as much as possible. Thus little additional overhead is required for the computation of multiple paths.

### A. Route Discovery

The route discovery process has two major phases: route request phase and route reply phase. The route discovery process computes the multiple loop free paths. The route discovery process will be initiated when a route is requested by a source node and there is no information about the route in its routing table. First, the source node generates an RREQ and then floods the packet to networks. The RREQ's are propagated to neighbours within the source's transmission range. They also broadcast the packets to their neighbours. The process is repeated until the destination receives the RREQ. When an intermediate node receives the RREQ, it performs the following process:

1. When an intermediate node receives the information of RREQ, either it sends the route reply if the node is destination, or it rebroadcasts the RREQ to it neighbours.
2. The node reads the information from the RREQ.

In order to transmit route reply packets to the source, the node builds a reverse path to the source. The node will insert the path to its multiple path lists. Otherwise, the node will ignore the path and discard the RREQ.

Link failures in ad hoc networks are caused by mobility, congestion, packet collisions, node failures, and so on. In the AOMDV protocol, the link layer feedback from IEEE 802.11 is utilized to detect link failures. If a node sends packets along the broken link, it will receive a link layer feedback. When a node detects a link break, it broadcasts route error (RERR) packets to its neighbours. The neighbours then rebroadcast the packets until all source nodes receive the packets. If a source node receives the RRER, it will remove every entry in its routing table that uses the broken link. Differing from single path routing protocols, the route error packets should contain the information not only about the broken primary path but also the broken backup paths. When the source node receives the RERR's, it removes all broken routing entries and uses the shortest backup paths as primary paths. The source node initiates a route discovery process where all backup paths are broken.

## IV. EAOMDV OVERVIEW

### A. Protocol Assumptions

In this section we present the operation of EAOMDV. The EAOMDV is ad-hoc Reactive routing protocol based on DSDV. EAOMDV is an extension of AOMDV, The goal behind the proposed protocol is to provide efficient recovery from "route failure" in dynamic network. To achieve this at the time of route discovery it computes the "received power of the receiver node" using the "link failure prediction technique". It calculates the received power to predict pre-emptively before the route failure.

In Ad Hoc networks route failure may occurs due to less received power, mobility, congestion and node failures. EAOMDV predict pre-emptively the route failure that occurs with the less received power.

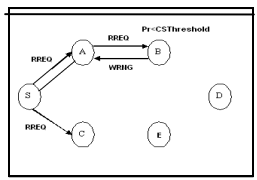

Figure 1: Route Request in EAOMDV

The above figure 1 shows the Route Request in EAOMDV; here the source node S wants to send the data to the destination D. Source S floods the RREQ packet to its neighbour nodes, Node A broadcasts the RREQ to the others nodes. After Broadcasting the Route Request Node B calculates the Receiver Power if it has less than Threshold it Drops the packet and sends the WRNG packet.

EAOMDV Protocol uses the main component:

1. A route update rule to establish and maintain stable multiple *loop-free* paths at each node using link failure prediction technique.

### B. Route Discovery

EAOMDV uses the Route Discovery method to discover the multiple node-disjoint paths. The route discovery uses the 2 different phases. Route Discovery phase and Route Reply phase. The Route Request phase is used to discover the path it broadcasts the RREQ packets. When an intermediate node receives the RREQ it performs following Steps.

**Step 1:** The node measures its Received Power (Pr) by comparing the "Carrier Sense Threshold".

**Step 2:** If Received power Pr of receiver node is less than threshold then the host drop the packet and send the warning packet.

**Step 3:** The routes with the less received power can be avoided, and then selects another path for transmission. The received power is goes on checking at every node.

The following algorithm 1 shows the Route update Rule for Route Request:

*Definations:*

> RREQ: A route request packet.
> CSThreshold: carrier sense threshold.
> Pr: Received power.
> WRNG: Warning Packet.
> **Procedure Route Request (RREQ) update**
> **Begin**
> When an intermediate node receives RREQ
>  if (Pr < CSThreshold) then
> Drop the packet and send the WRNG packet to source node.
> else if (Pr == CSThreshold)
> if ((Pr > CSThreshold)  then
> boardcast the RREQ to its neighbor nodes;
> endif
>  else
>  Drop the packet RREQ and choose the another path;
>  endif
> **End**

Algorithm 1. Route update rule for Route Request

Route Reply Phase, when the destination receives the route request packet, it sends route reply (RREP) packet to the source along the reverse paths created previously. The destination sends RREP to next nodes of reverse paths. They also forward the packet to next nodes until the source receives the RREP.

Whenever the Node Receives the RERR packet it immediately uses the Route maintenance phase to maintain the Route.

## V. SIMULATIONS AND RESULTS

### A. Simulation Environment

The EAOMDV was evaluated using the Ns-2 [9] simulator version 2.32 with CMU's multihop wireless extensions. In the simulation, the IEEE 802.11 distributed coordination function was used as the medium access control protocol. The physical radio characteristics of each wireless host were based on Lucent's WaveLan. WaveLan was direct spread spectrum radio and the channel had radio propagation range of 250 meters and capacity of 2Mb/sec. The AOMDV and enhanced AOMDV (EAOMDV) are to be compared in the simulation. Our results are based on the simulation of 50 wireless nodes forming an ad hoc network moving about in an area of 1500 meters by 300 meters for 100 seconds of simulated time. Nodes move according to the random waypoint model in a free space model.

### B. Simulation Results and Analysis

The following performance metrics used to evaluate the two routing protocols:

**Packet delivery ratio**: The ratio of the data packets delivered to the destination to those generated by the CBR sources.

**End to End delay**: Average time between data packets received by the destinations and data packet sent by CBR source. The data were collected only successfully delivered packets. The delay is determined by any factors such as buffering during route discovery, queuing at the interface queue and routing paths between two nodes.

**Overhead:** The number of routing packets transmitted per data packet delivered to the destination.

**Throughput:** the total size of data packets that are received in CBR destinations per second. It represents in whether the protocols make good use of network resources or not.

We report the results of the simulation experiments for the original AOMDV protocol with the EAOMDV. In this we analyze the performance metrics by the pause time.

Figure 2 compares the Average end-to-end delay by the different pause time. The routing protocols in varying in pause time. The graph demonstrates the node-disjoint EAOMDV performs better then AOMDV; End-to-End delay is less Because of less rout discovery.

Figure 3 plots the routing overhead of two routing protocols against pause time. Observe that node-disjoint EAOMDV has a less overhead than AOMDV. The reasons for less overhead is less route discoveries are initiated in EAOMDV, which lead to the flooding of RREQ.

Figure 4 compares the packet delivery ratio of routing protocols in varying pause time. In the simulation all the nodes move the same specified speed. The graph demonstrates the node-disjoint EAOMDV performs better then AOMDV; the paths in the Nod-Disjoint EAOMDV fail independently due to their node-disjoint property.

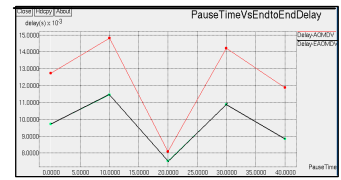

Figure 2: End-to-End Delay

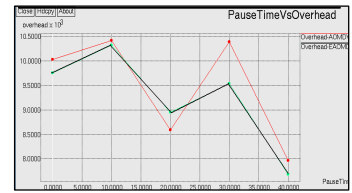

Figure 3: Overhead of AOMDV and EAOMDV

Figure 5 compares the throughput for both the protocols. Throughput of EAOMDV is better compared to AOMDV because of less Route Discovery; it saves the bandwidth and the network resources.

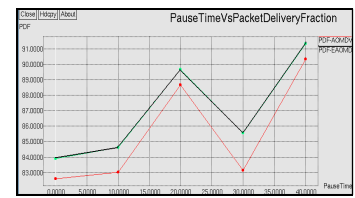

Figure 4: Packet Delivery Fraction

.

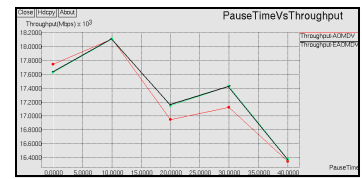

Figure 5: Throughput of AOMDV and EAOMDV

## VI. CONCLUSION AND FUTURE WORK

Multipath routing can be used in on-demand protocols to achieve faster and efficient recovery from route failures in highly dynamic Ad-hoc networks. In this project we have proposed an Enhanced Ad-Hoc on-Demand Multipath Distance Vector routing protocol that extends the AOMDV protocol to compute the stable path. There are three main contributions in this work. One is computing multiple loop-free paths at each node, another calculating received power of each node while transferring packets and node-disjoint route discovery mechanism.

Simulation results show that the throughput, packet delivery fraction EAOMDV is more than that of AOMDV. Also overhead of EAOMDV is less than that of AOMDV. This is because EAOMDV selects the node disjoint path pre-emptively before the path fails. If one path breaks then it selects an alternative and reliable node-Disjoint path. The advantage is less Route Discovery and Reduces the Route Error Packets.

In EAOMDV it uses multiple paths only one path is used at a time. It is possible to use multiple paths simultaneously for increasing data rate, which is not considered in this project. This aspect can be one area for future work.

ACKNOWLEDGMENT

This work is done in Poojya Doddappa Appa college of Engineering. I would like to express my gratitude to the supervisor of this work, professor .Sujata Terdal who gave me lot of valuable guidance and comments of this work.Special thanks to my family and friend without their support, I could not finish my Master degree studies.

AUTHORS PROFILE

I sujata V Mallapur. Completed pre high school and high school in Chetan School in Gulbarga. B.E and M.Tech in Poojya Doddappa Appa college of Engineering in the year of 2009. Presently working as lecturer in Appa Institute of Engineering And Technology, Gulbarga. My Interested Area is Ad-Hoc Network.